\documentclass[iop,apj,tighten]{emulateapj}

\usepackage{graphicx,natbib,amsmath,amssymb}
\bibpunct[, ]{(}{)}{;}{a}{}{,}
\newcommand{\simgt}{\hbox{\rlap{\raise 0.425ex\hbox{$>$}}\lower 0.65ex\hbox{$\sim$}}}
\newcommand{\simlt}{\hbox{\rlap{\raise 0.425ex\hbox{$<$}}\lower 0.65ex\hbox{$\sim$}}}

\slugcomment{Received 2010 April 15; accepted 2010 August 10; published 2010 September 23}

\shorttitle{STATISTICAL MECHANICS OF COLLISIONLESS ORBITS. I.}
\shortauthors{Hjorth \& Williams}

\begin{document}

\title{
%Statistical mechanics of stellar orbits -- I:
Statistical mechanics of collisionless orbits. I.
Origin of central cusps in dark-matter halos
%\thanks{Based on observations collected at the European th programme 10917.}
}


\author{ 
Jens Hjorth\altaffilmark{1}
and
Liliya L. R. Williams\altaffilmark{1,2,3}
}

\altaffiltext{1}{Dark Cosmology Centre, Niels Bohr Institute, 
University of Copenhagen, Juliane Maries Vej 30, DK-2100 Copenhagen \O, 
Denmark; jens@dark-cosmology.dk} 
\altaffiltext{2}{Department of Astronomy, University of Minnesota, 
116 Church Street SE, Minneapolis, MN 55455, USA; llrw@astro.umn.edu}
\altaffiltext{3}{Institute for Theoretical Physics, University of Z\"urich, 
Winterthurerstrasse 190, CH-8057 Z\"urich, Switzerland}


\begin{abstract}
We present an equilibrium statistical mechanical theory of collisionless 
self-gravitational systems with isotropic velocity distributions. Compared
to existing standard theories, we introduce two changes: 
(1) the number of possible microstates is computed in energy (orbit) space
rather than phase space and (2) low occupation numbers are treated more
appropriately than using Stirling's approximation. Combined, the two
modifications predict that the relaxed parts of collisionless 
self-gravitating systems, such as dark-matter halos, have a differential 
energy distribution $N(\varepsilon) \propto [\exp(\phi_0-\varepsilon) -1]$, 
dubbed ``DARKexp''. Such systems have central power-law density cusps 
$\rho(r) \propto r^{-1}$, which suggests a statistical mechanical origin of 
cusps in simulated dark-matter halos. 
\end{abstract}


%% subject header.  Objects should be in the appropriate "individual"
%% headers (e.g. quasars: individual, stars: individual, etc.) with the
%% additional provision that the total number of headers, including each
%% individual object, not exceed six.  The \objectname{} macro, and its
%% alias \object{}, is used to mark each object.  The macro takes the object
%% name as its primary argument.  This name will appear in the paper
%% and serve as the link's anchor in the electronic edition if the name
%% is recognized by the data centers.  The macro also takes an optional
%% argument in parentheses in cases where the data center identification
%% differs from what is to be printed in the paper.



\section{Introduction\label{introduction}}

The apparent universal light distributions in elliptical galaxies with two-body
collision relaxation time exceeding the age of the universe motivated 
\citet{1957SvA.....1..748O} and \citet{1967MNRAS.136..101L} 
to seek a fast relaxation process driving the systems toward equilibrium in 
a statistical mechanical sense. In a seminal paper, \citet{1967MNRAS.136..101L} 
introduced the process of `violent relaxation' (collective energy exchange 
between rapid potential fluctuations and individual particles) as responsible 
for a short time scale for relaxation. In the same paper, he introduced a new 
kind of statistical mechanics, that of distinguishable particles subject to 
an exclusion principle because collisionless dynamics precludes two parcels of 
phase space from being superimposed. In the non-degenerate limit the theory 
predicts that isothermal spheres are the maximum entropy equilibrium states of 
the process, as also found by \citet{1957SvA.....1..748O}.

The predictions of the theory are, however, not entirely satisfactory. It 
predicts infinite mass systems despite being derived under the constraints of 
fixed energy and mass.  It also predicts mass (phase-space density) segregation 
despite the dynamics being collisionless 
\citep[i.e., mass independent;][]{1986Ap&SS.122..299S,1993MNRAS.265..237H,2000ApJ...531..739N,2005MNRAS.361..385A,2005MNRAS.362..252A}.
The resulting isothermal sphere profile does not reproduce the light profiles 
of elliptical galaxies which was the original motivation. There is an 
arbitrariness in defining the initial states and whether to use a particle or 
phase element approach 
\citep{1978ApJ...225...83S,1987ApJ...316..497M,1987ApJ...316..502S,1993MNRAS.265..237H,1996ApJ...466L...1K,1997ApJ...484...58K,2008NewAR..52....1B}, 
which makes it difficult to assess whether the degenerate limit may be 
relevant. And finally, it is not obvious how to extend the statistical 
mechanical approach to (spherical) systems with anisotropic velocity 
distributions.

Some of these issues have been addressed in terms of modifications of the 
extent of the relaxation process, so-called 
incomplete violent relaxation \citep{1987MNRAS.229...61S}, 
relaxation in a finite volume \citep{1991MNRAS.253..703H}, 
or explicit scattering processes \citep{1992ApJ...397L..75S}.
Another approach has been to propose a change in the entropy functional to 
be optimized, applicable to non-extensive systems 
\citep{1988JSP....52..479T,1993PhLA..174..384P},
although this has been demonstrated not to work 
\citep{2007ApJ...655..847B,2008PhRvE..77b2106F}.

To address the infinite-mass problem, \citet{1996MNRAS.280.1089M}
\citep[and later][]{2006PhyA..367..269M,2008AdAst2008E...3D}
showed that appropriately dealing with small occupation numbers leads to 
finite-mass systems, similar to \citet{1966AJ.....71...64K} models, suitable 
for the description of globular clusters, which are driven by collisional 
relaxation.

The subject of the origin of universal collisionless self-gravitating
structure has gained renewed attention in recent years with the demonstration 
that the end products of numerical simulations of cosmological structure 
formation and dark-matter halos have remarkably universal profiles, 
in density as well as in pseudo phase-space density, $\rho/\sigma^3$. A new 
aspect, not foreseen by Lynden-Bell's theory or any modifications thereof, 
is that simulated dark-matter halos appear to have central density cusps, 
$\rho(r) \sim r^{-1}$, falling to $r^{-3}$ or $r^{-4}$, or even steeper, in 
the outer parts \citep{1997ApJ...490..493N,2004MNRAS.349.1039N,
2005ApJ...624L..85M,2010MNRAS.402...21N}. 
The origin of such cusps is unclear 
%\citep[see, e.g.,][]{2006ApJ...653..894H,2007ApJ...671.1147H,2009ApJ...690..102H}.
\citep[see, e.g.,][]{2006ApJ...653..894H}.
There also appears to be a relation between the local density slope and the 
degree of radial velocity anisotropy \citep{2006NewA...11..333H}.

In this paper, we explore a new avenue for implementing the collisionless 
nature of dark-matter halos. We suggest to partition state space in 
energy-per-unit mass shells rather than phase-space elements. Moreover, we 
implement the corresponding effect of small occupation numbers in this 
approach.  We show that the resulting state is a finite-mass system with a 
cusp, $r^{-1}$, falling in the outer parts to $r^{-4}$ for a completely 
relaxed, isolated, isotropic system. This captures the overall properties of 
simulated halos. In companion papers we compare in detail the resulting states 
with numerically generated systems suitable for testing our predictions.

\section{Statistical mechanics\label{sm}}

We start out, following \citet{1896gas.book.....B} and \citet{1957SvA.....1..748O},
by defining the number of possible states:
\begin{equation}
W = N! \prod_i \frac{ g_i^{n_i}}{ n_i!}
\end{equation}
\citep[as discussed in Section~4 we do not 
introduce an exclusion principle, cf.][]{1967MNRAS.136..101L,
1978ApJ...225...83S,1987MNRAS.229...61S}. 
Here, $n_i$ is the occupation number in cell of size $g_i$ in state space 
($\mu$ space). 

It is customary to take the continuous limit ($n_i \to n$), identify 
$\Gamma(n+1) = n!$, and optimize $\ln W$ under constraints of fixed total 
energy and total number of particles using a variational approach. This yields
\begin{equation}
\ln g - \psi(n+1)=\alpha+\beta E,
\label{var2}
\end{equation}
where 
$\psi(n)\equiv d \ln \Gamma / d n$ is the digamma function,
$\alpha$ and $\beta$ are Lagrange multipliers associated with the constraints 
of total number of particles and total energy, respectively, and 
$E=v^2/2+\Phi({\bf x})$ is the energy per unit mass, where 
$\Phi({\bf x})$ is the gravitational potential.

Next, one introduces Stirling's formula
\begin{equation}
\ln n! = \left ( n + \frac{1}{2} \right ) \ln n - n + \frac{1}{2} \ln (2\pi) + \frac{\theta(n)}{12 n}\ ; \ \ 0 < \theta(n) < 1
\end{equation}
and uses Stirling's approximation for large $n$
\begin{equation}
\ln n! = n \ln n - n \ ; \ \ n \gg 1
\label{stirling4}
\end{equation}
which implies
\begin{equation}
\psi(n+1)= \ln n\ ; \ \ n \gg 1.
\label{stirling5}
\end{equation}
This leads to
\begin{equation}
n = g \exp (-\alpha - \beta E).
\label{occupnum}
\end{equation}
This is the classical finding for the non-degenerate limit, as we did not
introduce an exclusion principle in Equation (1). In particular, identifying 
$\mu$ space with the 6-dimensional (${\bf x}$,${\bf v}$) phase space, the 
isothermal sphere is retrieved if phase space is divided up equally into 
equal-size phase-space cells $g$:
\begin{equation}
f({\bf x},{\bf v}) = A \exp \left (-\beta\left[\frac{1}{2} v^2+\Phi({\bf x})\right ]\right ).
\label{classicalf}
\end{equation}

Below we introduce two modifications to this standard approach, which, 
combined, give dramatically different structures, which turn out to be 
reminiscent of the end-products of numerical simulations of collisionless 
self-gravitating $N$-body systems.

\subsection{Orbit Space Versus Phase Space\label{orbits}}

The first major modification consists in noting that in an equilibrium 
collisionless system all particles retain their energies. This is not the 
case in an equilibrium collisional system, such as a classical gas, or a 
self-gravitating system dominated by two-body encounters, such as a globular cluster. 
The fundamental property of a collisionless system is that particles are 
distributed on orbits. 
In a relaxed system, once an energy is assigned to a particle it stays in a
restricted portion of phase space, an energy cell. For an isotropic system, 
energy is the only isolating integral. Hence, we argue that in partitioning 
state space, the fundamental partition is energy space, not phase space
\citep[see also][]{Efthymiopoulos07}.

This implies that the occupation number $n$, Equation~(\ref{occupnum}), should be 
interpreted as an indicator of the number of particles with a given energy, 
i.e.,
\begin{equation}
N(E) \propto \exp (-\beta E)
\label{DARKexp1}
\end{equation}
and not as the number of particles in a parcel of classical phase space, as 
is usually assumed.

\citet{1982MNRAS.200..951B}
\citep[see also, e.g.,][]{1990ApJ...356..359H,1992ApJ...397L..75S,1987gady.book.....B} 
found that elliptical galaxies obeying the $R^{1/4}$ law have energy 
distributions very similar to Equation~(\ref{DARKexp1}), however these authors' 
motivation for using this form was largely empirical. 

Given $N(E)$ one can, of course, recover the classical distribution function.
\citet{1987MNRAS.229...61S} point out that the occupation numbers in the 
classical phase-space can be related to those in the energy space if the 
former are assigned non-equal a priori weights, inversely proportional 
to the volume of phase space with a given energy, i.e.,
\begin{equation}
f(E) \propto g(E)^{-1} \exp (-\beta E),
\end{equation}
from which Equation~(\ref{DARKexp1}) is retrieved since 
$N(E)\equiv f(E)g(E)$ for isotropic systems. The relation between the 
phase-space density $f$ and the differential energy distribution $N$ is 
obtained assuming isotropy  from
$f({\bf x},{\bf v})d^3{\bf v} d^3{\bf x} = N(E) dE$ 
and the density of states $g(E) = d^3{\bf v} d^3{\bf x}/dE = 
16 \pi^2 \int_0^{r_{\rm max}(E)} \sqrt{2(E-\Phi)} r^2 dr$
\citep{1987gady.book.....B}. 

Another interesting aspect of Equation~(\ref{DARKexp1}), as pointed out by 
\citet{1982MNRAS.200..951B}, is that for systems with less mass at the centers 
than in the outer parts, $\beta$, and hence the effective temperature, 
$T=\beta^{-1}$, need to be negative \citep[see also][]{1989MNRAS.236..829M}. 
This is consistent with the self-gravitating nature of the systems, as these are 
characterized by negative heat capacity \citep{1987gady.book.....B}.

\subsection{Small occupation numbers\label{occupation}}

\subsubsection{Integer approach}

The second major modification is necessary because in systems with a finite
potential depth the occupation numbers in energy cells with very bound energies 
can become small (Equation~(\ref{DARKexp1})). A similar problem is encountered 
when the classical phase-space distribution function, $f(E)$, takes on an 
exponential form (Equation~(\ref{classicalf})), but in this case, because the 
temperature is positive, the small occupation numbers are found at the near 
escape energies, i.e., in the outer regions of systems. 
\citet{1996MNRAS.280.1089M} pointed out that in this case the Stirling 
approximation, Equation~(\ref{stirling4}), breaks down. Following 
\citet{1994AmJPh..62..515S}, he argued that the appropriate form for 
Equation~(\ref{occupnum}) is
\begin{equation}
n_i = [g_i \exp (-\alpha -\beta E_i)],
\label{occupnum2}
\end{equation}
where [$\cdot$] means rounding down to the nearest integer. 

Because the majority of particles in this latter case have energies near escape 
energies, the small occupation number modification has a dramatic effect on the
resulting structures. 
\citet{1996MNRAS.280.1089M} showed that Equation~(\ref{occupnum2}) introduces a 
cutoff which results in finite-mass systems, similar to the 
\citet{1966AJ.....71...64K} models of globular clusters. In effect,
\citet{1996MNRAS.280.1089M} analytically derived the well-known
\citet{1966AJ.....71...64K} models, which were originally obtained as a 
simple heuristic modification of the isothermal sphere's distribution function,
Equation~(\ref{classicalf}).\footnote{One can possibly improve upon 
\citet{1996MNRAS.280.1089M}'s Figure~1 by assigning different central potential 
values of the new models that match the old models. Then, the difference between 
the two will be confined to low density regions, where it belongs.}

For collisionless systems, where the temperature is negative, the cutoff 
occurs at energies close to that of the central potential value, i.e., close 
to the center of the system. Hence, the effect on the structures is not so 
dramatic in terms of total mass. But as we show below, it determines the inner 
density profile of the equilibrium systems.

\begin{figure}
\plotone{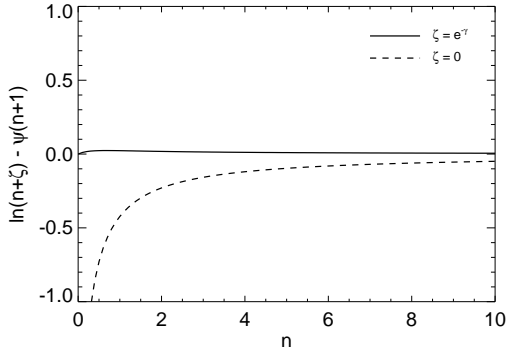}
\caption{
Approximations to $\psi(n+1)$. Comparison of the standard approach 
($\zeta=0$, dashed curve) and the approximation used here 
($\zeta=\exp(-\gamma)$, solid curve).
%This figure is also available as an mpeg
%animation in the electronic edition of the
%{\it Astrophysical Journal}.
}
\end{figure}

\subsubsection{Continuous approach}

Because a physical system is not expected to have a step-like differential 
energy distribution $N(E)$, a continuous version of Equation~(\ref{occupnum2}) 
is needed.
We do this by introducing a superior approximation to Stirling's formula which,
unlike Equation~(\ref{stirling5}), is not limited to large occupation numbers. 

We start by noting that when $n=0$, one has the exact result $\psi(1)=-\gamma$,
where $\gamma\approx 0.57721566...$ is Euler's constant. Combining this with 
the large $n$ limit, Equation~(\ref{stirling5}) leads to the approximation
\begin{equation}
\psi(n+1)\approx \ln(n + \zeta),
\end{equation}
with $\zeta= \exp(-\gamma) \approx 0.561459...$. While simple, this expression 
is a remarkably good approximation as shown in Figure~1 along with the 
classical large $n$ approximation (Equation~(\ref{stirling5}); $\zeta=0$). The 
deviation from the exact $\psi(n+1)$ is always positive, with a single maximum 
difference of 0.0237 at $n=0.680$. Using this in Equation~(\ref{var2}), one 
obtains a modification to Equation~(\ref{occupnum}),
\begin{equation}
n = g \exp (-\alpha-\beta E) - \zeta.
\end{equation}
Similar modifications have previously been proposed 
\citep{1954PNAS...40..149L,1993hjorth.thesis.H,2006PhyA..367..269M,2008AdAst2008E...3D}.

The Lagrange multiplier $\alpha$ is determined by requiring that $n=0$ at some 
energy $E'$. Thus, we get
\begin{equation}
n = \zeta (\exp (-\beta [E - E'])-1),
\label{occupnum3}
\end{equation}
thereby eliminating the cell size, $g$, which does not have a unique, 
physically meaningful value in gravitational systems. 

For a system which vanishes at the escape energy, $E' = 0$, and so the 
classical phase-space occupation number function is 
$n = \zeta (\exp (-\beta E)-1)$. The systems we are interested in vanish at a
finite central potential, $E'=\Phi_0$, and so the energy-space occupation 
number function becomes $n = \zeta (\exp (-\beta (E-\Phi_0))-1)$. This is the 
continuous version of Equation~(\ref{occupnum2}).

\subsection{DARKexp models}

Equation~(\ref{occupnum3}) with $E'=\Phi_0$ incorporates the two modifications we 
introduced in Sections~\ref{orbits} and \ref{occupation}. 
Identifying the occupation number $n$ as being proportional to $N$, the 
differential energy distribution, one obtains
$N(E) = A (\exp(-\beta [E-\Phi_0])-1)$,
where $A$ is determined by the mass of the system. In this expression, $E$ and 
$\Phi_0$ have units of energy, and $\beta$ is the inverse of temperature. We 
convert this to dimensionless form, using $\varepsilon=\beta E$ and 
$\phi_0=\beta\Phi_0$ (note that both $\varepsilon$ and $\phi_0$ are positive 
quantities for bound systems),
\begin{equation}
N(\varepsilon)=A(\exp[\phi_0-\varepsilon]-1).
\label{DARKexp}
\end{equation}
We dub this expression the DARKexp. It represents our prediction for 
fully relaxed, collisionless, self-gravitating, isotropic systems. Because
the arguments presented in this paper do not apply to non-isotropic systems, 
Equation~(\ref{DARKexp}) cannot be directly compared to the results of $N$-body
simulations. Having said that, we note that $N(\varepsilon)$ is not very
sensitive to anisotropy, as it depends primarily on the density profile and 
not on the dynamics of the system. 

The detailed structure of the DARKexp systems, including their density and 
velocity dispersion profiles, will be considered in an accompanying paper
\citep{WilliamsHjorth10}. In 
the next section, we discuss the limiting behavior of DARKexp models at small 
and large radii.

\section{Resulting structures\label{structures}}

\subsection{Limiting Power-Law Behavior at Small Radii\label{limit1}}

In a general spherically symmetric structure, the limiting form for the central 
potential can be assumed to be 
\begin{equation}
\phi=\phi_0-\phi_\alpha r^\alpha + \cdots
\end{equation}
Possion's equation $\nabla^2 \phi = 4 \pi G \rho$ yields a limiting power-law
behavior of the central density,
\begin{equation}
\rho(r)\propto r^{\alpha-2}.       %; \delta=2-\alpha
\end{equation}
For $\varepsilon \to \phi_0$ the distribution function then becomes
\begin{equation}
f(\varepsilon)\propto(\phi_0-\varepsilon)^{-(4+\alpha)/2\alpha}
\end{equation}
and the density of states becomes
\begin{equation}
g(\varepsilon)\propto(\phi_0-\varepsilon)^{(6+\alpha)/2\alpha}.
\end{equation}
For similar expressions, see \citet{1995MNRAS.276..679H},
\citet{2000ApJS..131...39W}, and
\citet{2004MNRAS.353...15A}. The differential mass distribution 
$N(\varepsilon)=f(\varepsilon)g(\varepsilon)$
then becomes 
\begin{equation}
N(\varepsilon)\propto(\phi_0-\varepsilon)^{1/\alpha}.
\end{equation}

In general, the value of $\alpha$ ranges from 0 for the singular isothermal 
sphere  to $\alpha=1$ for Navarro--Frenk--White or Hernquist profiles, to shallower slopes, 
reaching a flat core for $\alpha=2$.  Increasing $\alpha$ corresponds to 
increasingly steeper $\ln N(\varepsilon)$ versus $\varepsilon$ curves. For 
$\alpha\to\infty$, the system develops a hole in the central density profile, 
which is unphysical.

\subsection{Limiting Power-Law Behavior at Small Radii for DARKexp Models}

In the DARKexp model, Equation~(\ref{DARKexp}), 
$N(\varepsilon) \propto (\phi_0 - \varepsilon)$ as $\varepsilon \to \phi_0$. 
In other words, $\alpha = 1$ and we retrieve the central density cusps 
\begin{equation}
\rho(r) \propto r^{-1},
\end{equation}
known from numerical simulations. The corresponding distribution function is
\begin{equation}
f(\varepsilon) \propto (\phi_0-\varepsilon)^{-5/2}
%g(E) = (E-\Phi_0)^{7/2}.
\end{equation}
for $\varepsilon \to \phi_0$, which is the same as that of the 
\citet{1990ApJ...356..359H} model.

Thus our proposed differential energy distribution, the DARKexp, naturally 
predicts that the central density slopes of collisionless self-gravitating 
systems should asymptote to $-1$ at the centers of structures. Note that this 
slope is not the result of any specific dynamical process operating during the 
formation of halos, but a generic consequence of full relaxation.

The Appendix discusses the limiting behavior for more general cutoffs to 
$N(\varepsilon) \propto \exp(-\varepsilon)$.

\subsection{Limiting Behavior at Large Radii}

While we are focusing on the inner parts of halos in this paper, we note that
a full model for the entire system can be obtained by assuming that
$N(\varepsilon)$ is finite at the escape energy and zero above (as plotted in 
Figure~2). The rationale behind this is that during violent relaxation the 
escape energy is no special location and $N(\varepsilon)$ is expected to be 
continuous at what will eventually become the escape energy
\citep{1987IAUS..127..511J,1991MNRAS.253..703H}.
For such a model $f(\varepsilon) \propto \varepsilon^{5/2}$ and 
$\rho(r) \propto r^{-4}$ for $\varepsilon \to 0$ and $r \to \infty$, similar 
to the Hernquist model and broadly consistent with numerical simulations of 
dark-matter halos. However, if relaxation is not complete at radii where 
particles have near escape velocities, then the outer density profile slope 
may deviate from $-4$.

\begin{figure}
\plotone{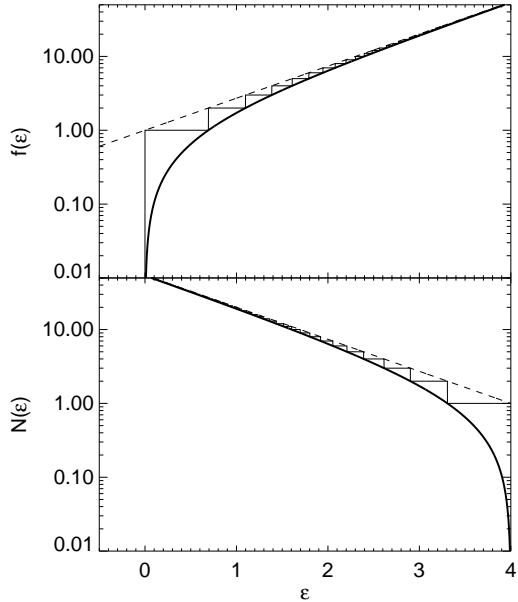}
\caption{Statistical mechanical predictions for $n$
(Equation~(\ref{occupnum3})). In the usual statistical mechanical approach, one
identifies $\mu$ space with ({\bf x},{\bf v}) phase space and $n$ with $f$
(upper panel). In the approach proposed in this paper, one identifies $\mu$
space with energy space and $n$ with $N$ (lower panel). We have assumed
$\phi_0 = 4$. Different line styles
signify different cutoffs: $\zeta = 0$ (no cutoff, thin dashed line),
$\zeta = \exp(- \gamma)$ (thick solid curve, DARKexp), and discrete approach
(thin solid line, Equation~(\ref{occupnum2})).
%This figure is also available as an mpeg
%animation in the electronic edition of the
%{\it Astrophysical Journal}.
}
\end{figure}

\section{Discussion}

\subsection{The approach to equilibrium}

In this paper, we used statistical mechanics to predict the energy distribution 
function of relaxed systems. A key feature of statistical mechanics approaches 
is that they deal only with the final equilibrium states, and derive these by 
equating them to the most probable  or maximum entropy states. We stress that 
our theory is therefore limited to the description of the final state of 
self-gravitating collisionless systems. Because the final state is an equilibrium 
state, it is the result of full relaxation and therefore does not rely on any 
additional assumptions.

Implicit in our derivation is the assumption of equal a priori 
probabilities in state space. While this can be accomplished through efficient 
mixing (ergodicity), our theory does not address specific physical mechanism
for attaining full mixing. In a companion paper \citep{WilliamsHjorth10},
we use the Extended Secondary 
Infall Model (ESIM) to test the DARKexp prediction, but stress that ESIM is a 
restricted physical model and is not equivalent to $N$-body simulations.

A straightforward way to evaluate the applicability of physical mechanisms for 
relaxation is to compare their end states. For example, the scattering model 
for violent relaxation introduced by \citet{1992ApJ...397L..75S} predicts
$N(E) \propto [(E-\Phi_0)^{-2}+C(E-\Phi_0)^{-1/2}]^{-1} \exp (-\beta E)$ 
which is clearly inconsistent with the DARKexp final state. On the other hand, 
the final state of ESIM halos is well fit with DARKexp.

\subsection{Collisional Versus Collisionless Systems}

Our prediction for $N(\varepsilon)$ applies to the end states of collisionless 
dynamics. Collisionlessness has several different consequences, many of which 
can be used in one's theory. \citet{1967MNRAS.136..101L} used the 
collisionless Boltzmann equation to introduce an exclusion principle to 
account for the incompressibility of the phase-space fluid. We use a different 
property; that collisionless dynamics implies the constancy of the energy per 
unit mass for each particle. Therefore, the collisionless nature of the 
problem is automatically included and there is no need for an exclusion 
principle.

In a system driven toward equilibrium by two-body relaxation, the situation is 
quite different. In this case, efficient relaxation implies that any particle 
can in principle end up anywhere in phase space and the usual partition of 
$\mu$ space is appropriate. In this case, taking into account low occupation 
numbers, one retrieves the \citet{1966AJ.....71...64K} $f(E)$ which is rounded 
down at near escape energies, while DARKexp's collisionless $N(E)$ is rounded 
down at highly bound energies. 

The density profiles of the two are clearly different, but both are very good 
approximations to what one sees in globular clusters and simulated dark-matter 
haloes, respectively.

\subsection{Anisotropy\label{anisotropy}}

In this paper we dealt exclusively with isotropic systems for which the 
differential energy distribution is a function of energy only. The full 
simulated structures, however, are definitely anisotropic, and exhibit a 
correlation between the density slope and anisotropy
\citep{2006NewA...11..333H}. 

In principle, our statistical mechanics formalism 
can be extended to spherical non-isotropic systems, and we plan to do so in 
a future publication. In a fully developed theory, the distribution function 
must depend explicitly on angular momentum, especially in the outer parts
\citep[e.g.,][]{1987MNRAS.229...61S}.  Whether the energy
distribution of the full theory will be significantly affected by the
introduction of angular momentum is yet unclear; numerical experiments
of \citet[][Figure~9]{2006ApJ...653...43M} seem to indicate that the DARKexp
form is still a good description of systems that underwent radial
orbit instability.

For now we note that even though the DARKexp prediction 
may not be expected to be an excellent fit to simulations at all energies, it 
should apply to the central regions of dark-matter halos, as these are 
isotropic.  Therefore, our explanation of the central density cusp of $-1$ 
applies to simulated halos.

\acknowledgments
The Dark Cosmology Centre is funded by the Danish National Research Foundation. 
L.L.R.W. is very grateful for the hospitality of the Dark Cosmology Centre at the University
of Copenhagen and the Institute for Theoretical Physics at the University of Z\"urich, 
where she spent the Fall of 2009 and the Spring of 2010, respectively.

%\appendix

\section*{Appendix}

\section*{Limiting Behavior at Small Radii for Different Exponential Cutoffs}

One might accept that $N(\varepsilon) \propto \exp(-\varepsilon)$ 
(Section~\ref{orbits}), but not the proposed form of the cutoff motivated by 
the low occupation numbers (Section~\ref{occupation}), for example, because our 
proposed transition from the integer to the continuous form is more algebraic 
than physical. In this section we address possible alternatives to the cutoff 
shape.

First, is a cutoff needed at all? If there is no cutoff, then the 
differential energy distribution is of the form Equation~(\ref{DARKexp1}), which 
would lead to a singularity in the central potential. To avoid that, one needs 
a cutoff.  One possible form is to simply truncate $N(\varepsilon)$ at some 
finite central value; $N(\varepsilon)=C$ at $\varepsilon=\phi_0$ and 
$N(\varepsilon)=0$ for $\varepsilon>\phi_0$. This would imply $\alpha=\infty$, 
and an infinite, positive central density slope, i.e., a hole 
(Section~\ref{limit1}) and divergent velocity dispersion, both of which are 
unphysical. 

Therefore, we conclude that a smooth cutoff is required, i.e., that at some
finite $\phi_0$, $N(\varepsilon=\phi_0)=0$. At energies near these highly bound 
energies, we can Taylor expand the non-truncated $N(\varepsilon)$. Using the 
abbreviation, $x=\phi_0-\varepsilon$: $\exp(x)\sim 1 + x + x^2/2 + \cdots$. 
Whatever the specific shape of the cutoff function, it too can be Taylor
expanded, and then subtracted from that of $\exp(x)$. Three outcomes
can result after this subtraction, depending on the exponent of $x$ of the 
surviving leading term, (1) $x^B$, where $B<1$  and $x$ and higher terms 
cancel out; (2) $x$; and (3) $x^2$, if the cutoff function goes exactly as 
$1+x$. In case (1) $\alpha>1$, and the central density slope will be 
shallower than $-1$. Case (2) leads to our main prediction of the central 
density slope of $-1$, while case (3) is the only way to get density slopes 
steeper than $-1$. In this case $\alpha=0.5$, and the central density slope 
is $-1.5$.

Since mathematically it is possible to get central density slopes shallower
or steeper than $-1$, it is ultimately the physical arguments that will dictate 
the form of the cutoff in $N(\varepsilon)$ and hence the slope value.

\bibliography{smo}

\begin{thebibliography}{45}
\expandafter\ifx\csname natexlab\endcsname\relax\def\natexlab#1{#1}\fi

\bibitem[{{Arad} {et~al.}(2004){Arad}, {Dekel}, \&
  {Klypin}}]{2004MNRAS.353...15A}
{Arad}, I., {Dekel}, A., \& {Klypin}, A. 2004, \mnras, 353, 15

\bibitem[{{Arad} \& {Johansson}(2005)}]{2005MNRAS.362..252A}
{Arad}, I., \& {Johansson}, P.~H. 2005, \mnras, 362, 252

\bibitem[{{Arad} \& {Lynden-Bell}(2005)}]{2005MNRAS.361..385A}
{Arad}, I., \& {Lynden-Bell}, D. 2005, \mnras, 361, 385

\bibitem[{{Barnes} {et~al.}(2007){Barnes}, {Williams}, {Babul}, \&
  {Dalcanton}}]{2007ApJ...655..847B}
{Barnes}, E.~I., {Williams}, L.~L.~R., {Babul}, A., \& {Dalcanton}, J.~J. 2007,
  \apj, 655, 847

\bibitem[{{Bindoni} \& {Secco}(2008)}]{2008NewAR..52....1B}
{Bindoni}, D., \& {Secco}, L. 2008, New Astron.\ Rev., 52, 1

\bibitem[{{Binney}(1982)}]{1982MNRAS.200..951B}
{Binney}, J. 1982, \mnras, 200, 951

\bibitem[{{Binney} \& {Tremaine}(1987)}]{1987gady.book.....B}
{Binney}, J., \& {Tremaine}, S. 1987, {Galactic Dynamics} (Princeton, NJ:
  Princeton Univ.\ Press)

\bibitem[{{Boltzmann}(1896)}]{1896gas.book.....B}
{Boltzmann}, L. 1896, {Vorlesungen \"uber Gastheorie, Vol.\ I} (Leipzig: Verlag
  von Johann Ambrosius Barth)

\bibitem[{{Dubey} {et~al.}(2008){Dubey}, {Menon}, {Pandey}, \&
  {Tripathi}}]{2008AdAst2008E...3D}
{Dubey}, R.~K., {Menon}, V.~J., {Pandey}, M.~K., \& {Tripathi}, D.~N. 2008,
  Adv.\ Astron., 2008, 1

\bibitem[{Efthymiopoulos {et~al.}(2007)Efthymiopoulos, Voglis, \&
  Kalapotharakos}]{Efthymiopoulos07}
Efthymiopoulos, C., Voglis, N., \& Kalapotharakos, C. 2007, in Topics in
  Gravitational Dynamics, ed. D.~Benest, C.~Froeschle, \& E.~Lega (Lecture
  Notes in Physics, Vol.~729; Springer: Berlin), 297

\bibitem[{{F{\'e}ron} \& {Hjorth}(2008)}]{2008PhRvE..77b2106F}
{F{\'e}ron}, C., \& {Hjorth}, J. 2008, \pre, 77, 022106

\bibitem[{{Hansen} \& {Moore}(2006)}]{2006NewA...11..333H}
{Hansen}, S.~H., \& {Moore}, B. 2006, New Astron., 11, 333

\bibitem[{{Henriksen}(2006)}]{2006ApJ...653..894H}
{Henriksen}, R.~N. 2006, \apj, 653, 894

\bibitem[{{Henriksen} \& {Widrow}(1995)}]{1995MNRAS.276..679H}
{Henriksen}, R.~N., \& {Widrow}, L.~M. 1995, \mnras, 276, 679

\bibitem[{{Hernquist}(1990)}]{1990ApJ...356..359H}
{Hernquist}, L. 1990, \apj, 356, 359

\bibitem[{{Hjorth}(1993)}]{1993hjorth.thesis.H}
{Hjorth}, J. 1993, {PhD thesis, Univ.~Aarhus}

\bibitem[{{Hjorth} \& {Madsen}(1991)}]{1991MNRAS.253..703H}
{Hjorth}, J., \& {Madsen}, J. 1991, \mnras, 253, 703

\bibitem[{{Hjorth} \& {Madsen}(1993)}]{1993MNRAS.265..237H}
---. 1993, \mnras, 265, 237

\bibitem[{{Jaffe}(1987)}]{1987IAUS..127..511J}
{Jaffe}, W. 1987, in Structure and Dynamics of Elliptical Galaxies, ed.
  {P.~T.~de Zeeuw}, IAU Symp.~127 (Cambridge: Cambridge Univ.~Press), 511

\bibitem[{{King}(1966)}]{1966AJ.....71...64K}
{King}, I.~R. 1966, \aj, 71, 64

\bibitem[{{Kull} {et~al.}(1996){Kull}, {Treumann}, \&
  {Boehringer}}]{1996ApJ...466L...1K}
{Kull}, A., {Treumann}, R.~A., \& {Boehringer}, H. 1996, \apjl, 466, L1

\bibitem[{{Kull} {et~al.}(1997){Kull}, {Treumann}, \&
  {Boehringer}}]{1997ApJ...484...58K}
---. 1997, \apj, 484, 58

\bibitem[{{Landsberg}(1954)}]{1954PNAS...40..149L}
{Landsberg}, P.~T. 1954, Proc.\ Natl.\ Acad.\ Sci., 40, 149

\bibitem[{{Lynden-Bell}(1967)}]{1967MNRAS.136..101L}
{Lynden-Bell}, D. 1967, \mnras, 136, 101

\bibitem[{{MacMillan} {et~al.}(2006){MacMillan}, {Widrow}, \&
  {Henriksen}}]{2006ApJ...653...43M}
{MacMillan}, J.~D., {Widrow}, L.~M., \& {Henriksen}, R.~N. 2006, \apj, 653, 43

\bibitem[{{Madsen}(1987)}]{1987ApJ...316..497M}
{Madsen}, J. 1987, \apj, 316, 497

\bibitem[{{Madsen}(1996)}]{1996MNRAS.280.1089M}
---. 1996, \mnras, 280, 1089

\bibitem[{{Menon} {et~al.}(2006){Menon}, {Dubey}, \&
  {Tripathi}}]{2006PhyA..367..269M}
{Menon}, V.~J., {Dubey}, R.~K., \& {Tripathi}, D.~N. 2006, Physica A, 367, 269

\bibitem[{{Merritt} {et~al.}(2005){Merritt}, {Navarro}, {Ludlow}, \&
  {Jenkins}}]{2005ApJ...624L..85M}
{Merritt}, D., {Navarro}, J.~F., {Ludlow}, A., \& {Jenkins}, A. 2005, \apjl,
  624, L85

\bibitem[{{Merritt} {et~al.}(1989){Merritt}, {Tremaine}, \&
  {Johnstone}}]{1989MNRAS.236..829M}
{Merritt}, D., {Tremaine}, S., \& {Johnstone}, D. 1989, \mnras, 236, 829

\bibitem[{{Nakamura}(2000)}]{2000ApJ...531..739N}
{Nakamura}, T.~K. 2000, \apj, 531, 739

\bibitem[{{Navarro} {et~al.}(1997){Navarro}, {Frenk}, \&
  {White}}]{1997ApJ...490..493N}
{Navarro}, J.~F., {Frenk}, C.~S., \& {White}, S.~D.~M. 1997, \apj, 490, 493

\bibitem[{{Navarro} {et~al.}(2004){Navarro}, {Hayashi}, {Power}, {Jenkins},
  {Frenk}, {White}, {Springel}, {Stadel}, \& {Quinn}}]{2004MNRAS.349.1039N}
{Navarro}, J.~F., {et~al.} 2004, \mnras, 349, 1039

\bibitem[{{Navarro} {et~al.}(2010){Navarro}, {Ludlow}, {Springel}, {Wang},
  {Vogelsberger}, {White}, {Jenkins}, {Frenk}, \&
  {Helmi}}]{2010MNRAS.402...21N}
---. 2010, \mnras, 402, 21

\bibitem[{{Ogorodnikov}(1957)}]{1957SvA.....1..748O}
{Ogorodnikov}, K.~F. 1957, SvA, 1, 748

\bibitem[{{Plastino} \& {Plastino}(1993)}]{1993PhLA..174..384P}
{Plastino}, A.~R., \& {Plastino}, A. 1993, Phys.\ Lett.\ A, 174, 384

\bibitem[{{Severne} \& {Luwel}(1986)}]{1986Ap&SS.122..299S}
{Severne}, G., \& {Luwel}, M. 1986, \apss, 122, 299

\bibitem[{{Shu}(1978)}]{1978ApJ...225...83S}
{Shu}, F.~H. 1978, \apj, 225, 83

\bibitem[{{Shu}(1987)}]{1987ApJ...316..502S}
---. 1987, \apj, 316, 502

\bibitem[{{Simons}(1994)}]{1994AmJPh..62..515S}
{Simons}, S. 1994, Am.\ J.\ Phys., 62, 515

\bibitem[{{Spergel} \& {Hernquist}(1992)}]{1992ApJ...397L..75S}
{Spergel}, D.~N., \& {Hernquist}, L. 1992, \apjl, 397, L75

\bibitem[{{Stiavelli} \& {Bertin}(1987)}]{1987MNRAS.229...61S}
{Stiavelli}, M., \& {Bertin}, G. 1987, \mnras, 229, 61

\bibitem[{{Tsallis}(1988)}]{1988JSP....52..479T}
{Tsallis}, C. 1988, J. Stat.\ Phys., 52, 479

\bibitem[{{Widrow}(2000)}]{2000ApJS..131...39W}
{Widrow}, L.~M. 2000, \apjs, 131, 39

\bibitem[{{Williams} \& {Hjorth}(2010)}]{WilliamsHjorth10}
{Williams}, L.~L.~R., \& {Hjorth}, J. 2010, \apj, 722, 856

\end{thebibliography}

\clearpage

%% the file, uncomment the \plotone, \plottwo, and
%% If you need a layout that cannot be achieved with \plotone or
%% \plottwo, you can invoke the graphicx package directly with the
%% \includegraphics command or use \plotfiddle. For more information,
%% please see the tutorial on "Using Electronic Art with AASTeX" in the
%% documentation section at the AASTeX Web site,
%% http://www.journals.uchicago.edu/AAS/AASTeX.
%%
%% The examples below also include sample markup for submission of
%% supplemental electronic materials. As always, be sure to check
%% the instructions to authors for the journal you are submitting to
%% for specific submissions guidelines as they vary from

%% This example uses \plotone to include an EPS file scaled to
%% 80% of its natural size with \epsscale. Its caption
%% has been written to indicate that additional figure parts will be
%% available in the electronic journal.

\clearpage

\clearpage

\end{document}